# ORBITAL AND PLANETARY CHALLENGES FOR HUMAN MARS EXPLORATION

Malaya Kumar Biswal M[*], Noor Basanta Das[†], and Ramesh Naidu Annavarapu[‡]

The space challenges do exist at every stride on a human expedition to Mars that arise due to galactic natural phenomena and artificial technologies. This paper emphasizes on Mars orbital and planetary challenges encountered from orbit to the surface exploration. The Mars orbital challenges embrace hazards of cosmic radiation and asteroid impact in orbit, disrupted communication relay from the ground, the intelligence of planetary weather clearance, and execution of successful entry, descent, and landing. Comparably planetary challenge encompasses identifying scientific landing site, an intrusion of erratic environment and weather, complexity in in-situ resource extraction and exploitation, navigation and surface mobilization, and retarded communication from relay orbiters. The prime intent of this study is to present every prospective challenge and its recommendations impending human settlement on Mars.

**INTRODUCTION**

The human race originated from Africa, but curiosity and necessity did not let us stay there for long. Over millions of years, humans sought out more resources and ventured across the continent on foot and then out of it over the seas by building boats and ships to lands they never knew existed. This presented them with an inexhaustible amount of problems like starvation, lack of shelter, predators, carrying sufficient resources etc. Our ancestors overcame these challenges over time through inventions and innovations towards the then technology or tools available for land exploration. This led to an era of extensive overseas exploration by the Europeans in the early 15th century dubbed as the "Age of Discovery". The history of space exploration has been no different. As curiosity would have it, since the ancient times humans have looked up to the sky in wonder, what is out here? Is it possible to go there? Is anyone else out there? The ancient Egyptians and Greeks studied the stars and their movements as did the Indians. But it was the launch of the Sputnik 1 in 1957 that marked the beginning of the "Space Age". Challenges like spacecraft launch, flight path, overcoming gravity of the earth and the lack thereof in space, and human error have persisted since the first space mission. With recent advancements in space technology like lighter and better materials for spacecraft, tracking systems, better communication into and from space pose a major challenge to human exploration beyond LEO and we have overcome a lot of these threats to a successful mission into space (References 1 and 2).

---

[*] Graduate Researcher, Department of Physics, Pondicherry University, R.V, Nagar, Kalapet, Puducherry – 605 014, India.
[†] Graduate Researcher, Department of Physics, Pondicherry University, R.V, Nagar, Kalapet, Puducherry – 605 014, India.
[‡] Associate Professor, Department of Physics, Pondicherry University, R.V, Nagar, Kalapet, Puducherry – 605 014, India.
**Contact:** malaykumar1997@gmail.com; mkumar97.res@pondiuni.edu.in. (**Malaya Kumar Biswal**); noorbasantadas@gmail.com (**Noor Basanta Das**); and rameshnaidu.phy@pondiuni.edu.in; arameshnaidu@gmail.com (**Dr. A. Ramesh Naidu**).

At first glance, Mars does not seem very different from Earth, with moons, polar ice caps, large valleys, liquid water under its surface and a day slightly longer than ours. But unfortunately, a human mission to the red planet poses a number of threats like cosmic radiation, descent and landing, identifying landing sites, isolation and confinement etc. These threats can be mainly divided into two types, namely Mars orbital Challenges and Planetary surface challenges. In this paper, we enumerate and detail all the possible problems in both categories and try to come up with possible solutions for them. Overall outline map for orbital and planetary challenges is shown in Figure 1.

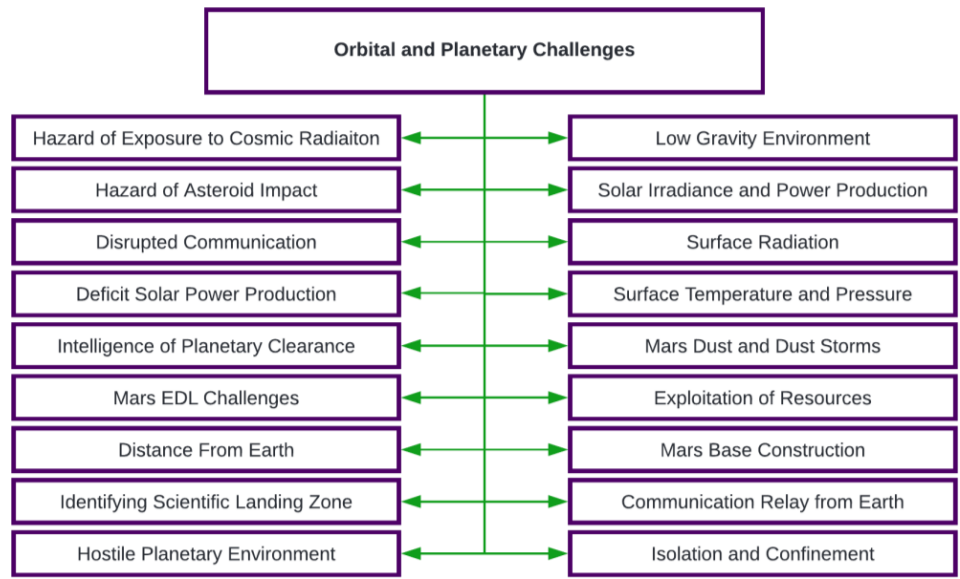

**Figure 1 Outline Map for Orbital and Planetary Challenges for Human Mars Missions**

## MARS ORBITAL CHALLENGES

Mars orbital challenges are the various obstacles that a planned mission to Mars will face and need to overcome on its journey from launch off the surface of Earth to the orbit of Mars. These obstacles can compromise or make things more difficult for a successful mission at any point on its way to the destination if not taken into account and prepared for in advance. The Mars-bound challenges embrace hazards of cosmic radiation and asteroid impact in orbit, disrupted communication relay from the ground, the intelligence of planetary weather clearance, and execution of successful entry, descent, and landing. We will discuss one by one in detail from the next section.

## HAZARDS OF EXPOSURE TO COSMIC RADIATION

The first and the most menacing of all the hazards a mission to deep space could face is radiation. Radiation becomes a concerning problem as soon as the spacecraft leaves the protective layers of Earth's atmosphere. Consequent to successful Mars Orbital Insertion, the Mars-bound challenges commence, as the crewed space vehicle need to strand in Mars orbit until further instructions from the ground for their next move. Because the spaceship undergoes preliminary checks and validation of its components and communication relay system before Mars atmospheric entry. So, during their stay in Mars orbit, the astronauts along with their spacecraft are exposed to high dosage of harmful galactic and extragalactic cosmic radiation as compared to low earth orbit. This radiation levels range from a minimum of 1.07 millisieverts per day to a maximum of 1.4 millisieverts per day ultimately increasing the possibility of prolonged cancer and related diseases (References 3 and 4). Hence, astronaut being sheltered under the Martian environment (on the surface of Mars) seems safer than stranding in low Mars orbit (LMO). So we recommend grounding the whole habitat module to the planetary surface from the perspective of crew health and safety. Comparison of radiation exposure in Earth and Mars orbit is shown in Figure 2[***].

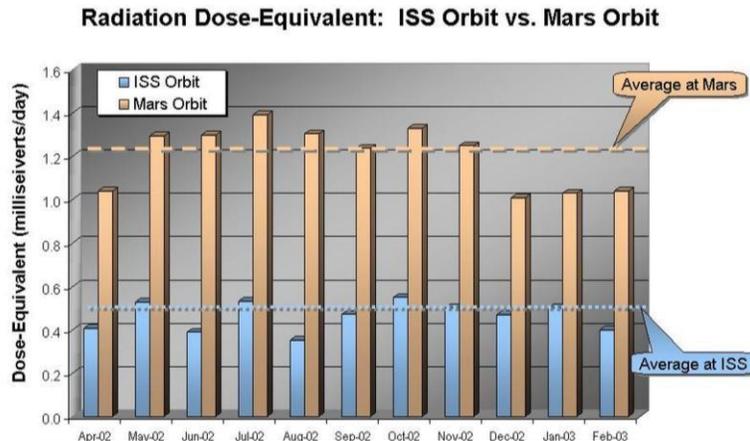

**Figure 2 Comparison of Radiation Dose Equivalent – Earth (ISS) and Mars Orbit**

**HAZARD OF ASTEROID AND METEOROID IMPACT**

Astronauts aboard space vehicle in orbit or spacecraft orbiting the red planet are vulnerable to the hazard of an asteroid impact and capable of damaging the spacecraft components. The asteroids sizes vary from micro to macro asteroids and are ejected from the main asteroid belt due to the probabilistic collisional events occurring at a distance ranging from 2.4 to 3.4 AU from the Sun with a relative velocity of approximately 8.0 km/sec (Reference 5). Larger asteroids can be mitigated or destroyed by directing into the Mars atmosphere, but the problem is with micrometeoroid or micro-asteroid which are travelling at approximately of more than 10km/sec can cause severe damage to the spacecraft component. The damages include rupture of spacecraft fuel tank through colliding and penetration, damage of astronaut's spacesuit during extravehicular activity, and depleting the solar arrays affecting the power production. However, this challenge cannot be completely eradicated, but proper attention is required while fabricating the sensitive component of space vehicles (i.e., fuel tank, solar array, and glasses of life-support systems) before initiating the interplanetary exploration missions (sufficient thickness of walls of the fuel tank needs to be considered during fabrication) and before stepping out for EVA. The challenge of asteroid impact will continue to exist for future interplanetary missions beyond Mars as the main asteroid belt lies between Mars and Jupiter and spreads up to hundreds of kilometers (approximately at 2.4 AU to 3.4 AU from the Sun) in interplanetary space with asteroid size ranges from 50 to 150 km in diameter (References 6 and 7).

**DISRUPTED COMMUNICATION**

Human exploration missions to Mars, as well as Deep Space predominantly require advanced communication systems to guide and land the spacecraft modules safely on Mars or any other planetary surface. Because a small misstep during EDL phases my cost mission tragedy as two-way communication interlinks span about 20 to 45 minutes and unreachable throughout the solar synodic period or solar conjunctions. Hence advanced communication system with laser-guided systems or via Mars Telecommunication Orbiter (MTO) or real-time decisiveness is ideally recommended for crewed missions and safer EDL performance (References 8 and 9).

**DEFICIT SOLAR POWER PRODUCTION**

Power production at Mars appears to be the most strenuous task for Martian satellites since the intensity of solar irradiance evanesces from Earth to Mars. Hence for a crewed missions thousands of watts are essentially required. So larger solar arrays capable of outstretching their solar cells are preferred to meet the energy requirements to power the space vehicle and estimated power production rate is 100 watts per square meter (at a solar intensity of 588 W/m2). Further, this power option is limited during the Mars solar conjunctions when Earth and Mars are far from each other having Sun at the median point for a period of 10-15 days (Reference 10). Therefore, we can alternatively exploit radioisotope thermoelectric generators (RTG) to afford the basic power necessities thereby mounting the RTG at a safe distance from the crewed module with proper shielding (to avoid the effects of nuclear radiation) (Reference 11).

**PLANETARY CLEARANCE AND EXECUTION MARS ATMOSPHERIC ENTRY**

Mars Entry, Descent, and Landing is a sturdy assignment for planetary landers due to the limitation of uncertainty in predicting the natural hindrance (prevalence of the variable environmental condition and dust storms) and EDL technology. Despite this challenge, Mars entry may impair the communication system and can cause damage to the landing module. Therefore, these issues can be eliminated by reporting the crews in advance who are grasped in Mars orbit about the environmental condition forecasted by Mars relay orbiters and ground rovers. It ensures crew about the planetary clearance and their next move. Further, the crewed landers are expected to perform orbital entry instead of direct entry which is considered as the safest approach and recommended for crewed landing (Reference 12). Because it reduces the entry velocity for increased ballistic coefficient and faster aerocapture and also it reduces the risk of crash landing due to limited EDL period, and design flexibility. Further, it enables the crews and their landing modules for an effective preparation to perform safer Mars entry. We have not reviewed much about Mars EDL challenges as it was technically reported by Braun (Reference 13).

**DISTANCE FROM THE EARTH**

The most apparent challenge is the vast distance between Earth and Mars. On average the distance is 225 million kilometers. The trip will take roughly three years compared to the 3 days trip to the Moon. Unlike the team at the ISS in case of medical events or any emergencies, the Mars mission cannot reach out for help to Earth. The mission cannot be resupplied once it leaves the low Earth orbit, therefore planning and self-sufficiency are essential keys to a successful manned mission. With a one-way delay of up to 20 minutes in communications, the astronauts should be trained well enough to confront an array of situations without any support from their team on Earth. We do not have any logistics or data due to a lack of prior missions which makes predicting any hazards on the way extremely difficult (Reference 14).

**PLANETARY SURFACE CHALLENGES**

Planetary surface challenges are the obstacles that a human mission to Mars will face in the second phase of its mission that is building an outpost for human settlement. Planetary challenges encompass identifying scientific landing site, an intrusion of erratic environment and weather, complexity in in-situ resource extraction and exploitation, navigation and surface mobilization, and retarded communication from relay orbiters (Reference 15).

## IDENTIFYING SCIENTIFIC LANDING ZONE

Exploration zone with good scientific interest and resource determines the success and sustainability of the mission. The scientific site should meet all necessities for the crews and should have affordable native resources for exploitation to keep alive the crew during extended surface stay mission. NASA has identified forty-seven candidate landing site for robotic and manned exploration (References 16 and 17). Of these Meridiani Planum seems to be the best site for first Crewed Mars landing and Base establishment. Because, it holds an ideal site for promising resources which includes the potential for water exploitation, raw materials for infrastructure and construction purposes, and some valuable minerals. Meridiani Planum is located at 50°N and 50°S with an elevation of below +2 km (MOLA). Additionally, it enables crews for practicing planetary cropping and plantation, food production with efficient solar power production as it lies near-equatorial latitude, and facilitates for accomplishing multidisciplinary scientific goals in terms of atmosphere, astrobiology and geosciences. The additional feature of Meridiani Planum was reviewed by Clarke (Reference 18).

## HOSTILE PLANETARY ENVIRONMENT

### Density of Mars Atmosphere

Due to the thin atmosphere of Mars, the planet is incapable of shielding its surface from being exposed to harmful cosmic radiation and pose a threat to the living astronaut on the surface. Similarly, its lean atmosphere with lower density forbids lander modules from faster aerocapture thereby limiting the EDL period (Reference 13). In addition to this, the composition of Mars atmosphere $CO_2$ (95%) and $O_2$ (0.17%) stands a challenge for the astronaut to breathe outside their confined habitat or spacesuit (Reference 19).

### Low-Gravity Environment

Astronauts on Mars gets exposed to the low gravity environment affecting the periodic pattern of heartbeat, blood flow rate, reduction in bone density of astronaut and weaken muscles, and physical movements. The human body takes time to adapt their internal organs to sustain their presence under low gravity environment. Hence, these issues can be managed by frequent practicing of physiotherapy and physical exercises (References 20, 21 and 22).

### Solar Irradiance and Power Production

The challenges of solar power do exist at every extremity beyond LEO. For a manned mission, this complication comes during the interplanetary voyage, stranded in Mars orbit, and on Mars surface. However, the intensity of solar irradiance weakens from orbit to the surface and the mean power production rate is about 20 watts per square meter (Source: InSight Mars Lander) (Reference 23). Hence, the structure of extendable solar arrays can be employed for mass electricity production but instead, the nuclear thermoelectric generators (NTG) will be the ideal choice for power source on the surface during the day and night. Furthermore, the intensity of solar irradiance influences the environmental temperature that poses a challenge against manned Mars exploration to stabilize the thermal stability of crewed habitat (Reference 24 and 25).

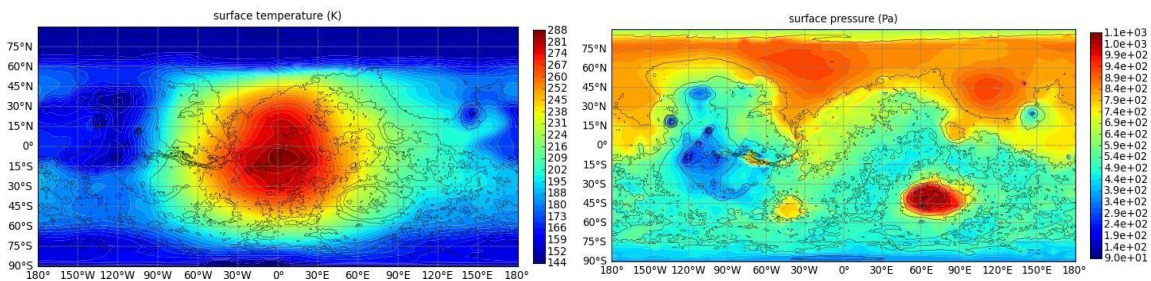

**Figure 3 Distribution of Surface Temperature and Pressure**

**Temperature and Pressure**

The frequency of temperature on Mars varies for every Martian year. Findings and observations from diverse spacecraft have shown that the temperature variance ranges from 120K near poles to 293K at the equator with an average of -210K (Reference 26). Hence at this range of temperature, the crew may experience complication in maintaining the thermal stability of both habitats and their body's internal temperature to keep them warm against hypothermia and its related health effects (Reference 27). Contradictory to temperature the pressure varies from 400 Pa to 870 Pa in accordance with the seasonal pattern and pose a greater challenge to maintain a pressurized environment. So, this challenge can be addressed by deploying a Mars sub-surface habitat to balance the thermal stability during the day and night (Reference 28). Distribution of surface temperature and pressure is shown in Figure 3[*].

**Surface Radiation**

Unlike Earth, Mars does not have an extensive magnetosphere, dense atmosphere or an ozone layer (Reference 29). Hence, half of all radiation received from the surface reaches the ground. As measured by the Radiation Assessment Detector (RAD) aboard the Curiosity rover the absorbed dose and dose equivalent from galactic cosmic rays and solar energetic particles on the surface of Mars show an average GCR dose equivalent rate of 0.67 millisieverts per day from August 2012 to June 2013[†]. A 500-day mission on the surface would bring total exposure to around 1mSv. Three years on the surface of Mars exceeds the radiation dose limit on NASA astronauts throughout their entire career. To avoid excess radiation exposure, the habitats can be shielded with a thick layer of $CO_2$ which can be harvested directly from the atmosphere. This layer of dry ice can be further covered with a layer of dirt to increase the level of protection[‡§] (Reference 30).

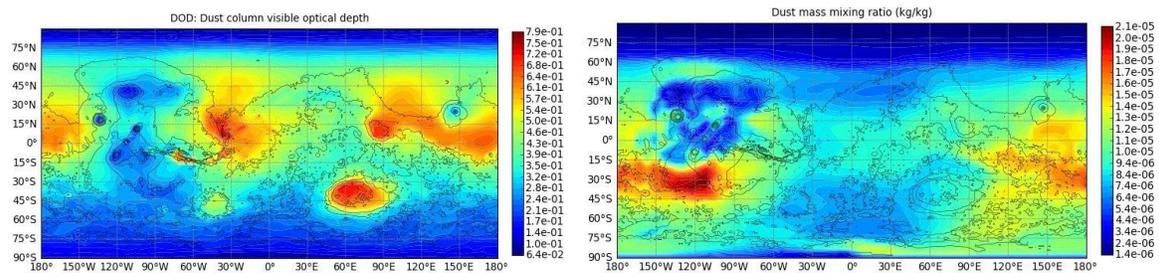

**Figure 4 Dust Column and Dust mass mixing ratio on Mars[**]**

---

**Mars Dust and Dust Storms**

Mars dust is very small and fine compared to Earth. They are also slightly electrostatically charged which makes them stick to everything. They can stick to the spacesuits, machinery, and even solar panels decreasing the amount of sunlight that hits the panels. This can further reduce the energy produced by solar panels. Global storms can also present a secondary issue, throwing enough dust into the atmosphere to reduce sunlight reaching the surface of Mars. Engineers need to take this into account while designing equipment for Mars missions. Mars dust contains a very high amount of toxic perchlorate salts. The habitats need to Spacesuits need to designed in ways that they never carry in any dirt into the habitats[*]. Distribution of Dust effective radius and dust deposition on flat surface is shown in Figure 5.

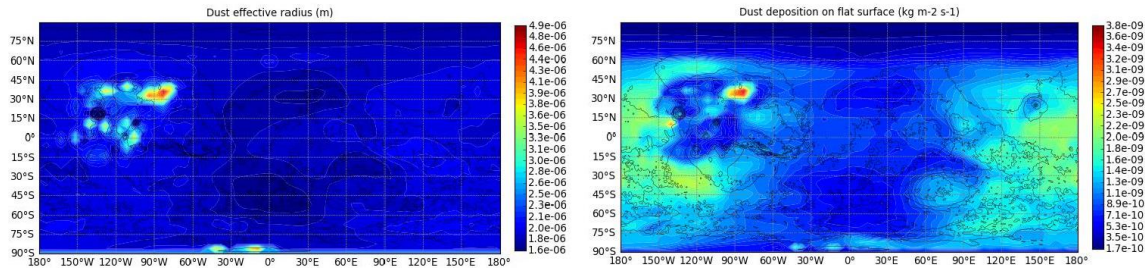

**Figure 5 Distribution of Dust Effective Radius and Dust Deposition on Mars**[†]

# EXPLOITATION OF RESOURCES

The challenge associated with the exploitation of resources is locating a robust site for exploration as well as extraction. Because the distinct site is associated with divergent distribution and concentration of resources, the form at which they exist, and the quantity of contaminants from the aspect of planetary protection[‡]. Since transportation of resources from different sites to the base is limited due to the constraints in surface mobility and lack of long-range rovers. Further, the unavailability of the testbed to demonstrate and validate ISRU instruments under a critical and low gravity environment poses a technical challenge on the surface (References 31). Furthermore, the uncertainty in system reliability and integration remains the greatest challenge at the very beginning of the Mars Base foundation. Current NASA plans for the future Mars In-Situ Resource Utilization is briefly presented by G. Sanders (Reference 32).

# CONSTRUCTION OF MARS BASE AND SURFACE MOBILIZATION

For a limited crew member at the initial stage of colonization, the habitats and the other modules can be exported from the Earth. But for a larger number of the population, a Mars Base is required and construction of this massive base using labour force is not obvious due to the vulnerability of Martian Environment and exhaustion of limited survival resources. Hence robotic based construction is beneficial from the perspective of crew health and also in retaining survival resources. Similarly, system reliability and its extended operation in a critical environment remain inconsistency due to technical challenges such as unproven technologies in the appropriate testbed and solar power deficiency. In addition to this, base construction is supported by load transportation systems from various resource mining sites. This mode of transportation may prompt the systematic servicing and repairing of vehicles (References 33,34 and 35).

---

[*] The Fact and Fiction of Martian Dust Storms. Accessed from https://www.nasa.gov/feature/goddard/the-fact-and-fiction-of-martian-dust-storms on 30 November 2020.
[†] Madeleine, J. B., Forget, F., Millour, E., Montabone, L., & Wolff, M. J. (2011). Revisiting the radiative impact of dust on Mars using the LMD Global Climate Model. Journal of Geophysical Research: Planets, 116(E11).
[‡] Toxic Mars: Astronauts Must Deal with Perchlorate on the Red Planet. Accessed from https://www.space.com/21554-mars-toxic-perchlorate-chemicals.html on 30 November 2020.

**COMMUNICATION RELAY FROM EARTH**

The communication interlink from Earth to Mars is significant at the initial phase of human civilization. Since the vast red planet is a completely new and inexperienced environment which is far different from the lunar mission experience (Apollo Missions) (Reference 36). So the astronaut needs to stay tethered and get updated about their mission operations and next move. Communication systems also play a major role in navigating astronauts to enable surface mobility thereby reporting the map and exploration zones. These modes of communication systems are interrupted or remain unavailable during the solar conjunctions period for 10-15 days. Hence, this challenge addressed by establishing efficient network wither via serious of Mars Telecommunication orbiter constellation network or parking the communication satellites in non-kaplerian orbits (Reference 37).

**ISOLATION AND CONFINEMENT**

Human beings need to have continuous contact and communication with other humans. Isolation and confinement can become a key issue regarding the mental health of our crew. Sleep loss, circadian desynchronization, and work overload compound this issue and may lead to performance decrements, adverse health outcomes, and compromised mission objectives. To address this, issue our crews need to be carefully chosen through intense psychological screening, trained to be compatible and work together for years in space[*].

**CONCLUSION**

A human mission to Mars and building a permanent base will be the toughest challenge we have ever faced as a species, but with proper planning, resources, training and advancement in technology we can overcome them. So, we have highlighted significant challenges and provided a guide to the various hazards of Human Mars exploration and their possible solutions. This will be greatly helpful to the global space communities in planning future missions and the consequent habitation of Mars. Interplanetary Challenges and overall Mars expedition challenges were technically reviewed by M.K. Biswal (References 15 and 38).

**ACKNOWLEDGEMENTS**

The first and second author would like to thank their supervisor, **Prof. A. Ramesh Naidu. Ph.D** (author), for the patient guidance, encouragement and advice he has provided throughout their time as his student. Further, we would like to extend our sincere thankfulness to all of our lovable friends for their financial support for the conference participation.**CONFLICT OF INTEREST**

The authors have no conflict of interest to report.

---

[*] 5 Hazards of Human Spaceflight. Accessed from https://www.nasa.gov/hrp/5-hazards-of-human-spaceflight on 30 November 2020.

## ACRONYMS

| | | |
|---|---|---|
| EDL | - | Entry, Descent and Landing |
| EMU | - | Extra-Vehicular Maneuvering Unit |
| GCR | - | Galactic Cosmic Radiation |
| HMO | - | High Mars Orbit |
| ISRU | - | In-Situ Resource Utilization |
| ISS | - | International Space Station |
| LEO | - | Low Earth Orbit |
| MOI | - | Mars Orbital Insertion |
| MOLA | - | Mars Orbiter Laser Altimeter |
| MRO | - | Mars Reconnaissance Orbiter |
| MTO | - | Mars Telecommunication Orbiter |
| NASA | - | National Aeronautics and Space Administration |
| NTG | - | Nuclear Thermoelectric Generator |
| RAD | - | Radiation Assessment Detector |
| RTG | - | Radioisotope Thermoelectric Generator |
| SCR | - | Solar Cosmic Rays |